\documentclass[aps,pre,twocolumn,showkeys,floatfix]{revtex4}%
\usepackage{graphicx}
\usepackage{epsf}
\usepackage{amsmath}%
\usepackage{amsfonts}%
\usepackage{amssymb}

\begin{document}

\title{Static Pairwise Annihilation in Complex Networks}

\author{M. F. Laguna$^{1,2}$}

\author{M. Aldana$^{1,3}$}
\email{max@fis.unam.mx}

\author{H. Larralde$^{3}$}

\author{P. E. Parris$^{1,4}$}

\author{V. M. Kenkre$^{1}$}

\affiliation{(1) Consortium of the Americas for Interdisciplinary
Science, University of New Mexico, Albuquerque, USA. }

\affiliation{(2) Centro At\'{o}mico Bariloche, CONICET and Instituto
Balseiro, San Carlos de Bariloche, Argentina. } 

\affiliation{(3) Centro de Ciencias F\'\i sicas-UNAM, Cuernavaca,
M\'exico.}

\affiliation{(4) University of Missouri-Rolla, Rolla, Missouri, USA.}

\date{\today}

\begin{abstract}
We study static annihilation on complex networks, in which pairs of
connected particles annihilate at a constant rate during time.
Through a mean-field formalism, we compute the temporal evolution of
the distribution of surviving sites with an arbitrary number of
connections. This general formalism, which is exact for disordered
networks, is applied to Kronecker, Erd\"os-R\'enyi (i.e. Poisson) and
scale-free networks.  We compare our theoretical results with
extensive numerical simulations obtaining excellent
agreement. Although the mean-field approach applies in an exact way
neither to ordered lattices nor to small-world networks, it
qualitatively describes the annihilation dynamics in such
structures. Our results indicate that the higher the connectivity of a
given network element, the faster it annihilates. This fact has
dramatic consequences in scale-free networks, for which, once the
``hubs'' have been annihilated, the network disintegrates and only
isolated sites are left.
\end{abstract}



\maketitle

\section{Introduction}
\label{sec:introduction}

The reaction dynamics of particles that undergo pairwise mutual
annihilation is of interest to various fields of physics, including
diffusion controlled reactions \cite{redner1,redner2}, and exciton
annihilation in molecular crystals \cite{molec}. In these examples the
particles are free to move or diffuse throughout the system, and when
two complementary particles collide, they annihilate.  This kind of
dynamics can be modeled by adopting a network approach in which the
particles, rather than moving across the system, are placed at the
vertices of a given network. The edges of the network would represent
direct interactions between the particles at a given time, or
interactions that might occur during time as the particles diffuse
throughout the system.  This approach has proven to be useful in
understanding the collective dynamical behavior in many-particle
systems \cite{vicsek,selfdriven}.  In some other cases, the
annihilating particles can be accurately represented as being fixed on
a lattice.  As an example, consider the pycnonuclear reactions taking
place in the interior of dense stars \cite{pyc1,pyc2,pyc3}. In this
case, the atom nuclei undergoing mutual annihilation become frozen
into a regular crystalline lattice structure due to the high densities
of the cores. Other examples are the irreversible sequential
adsorption and ``car parking'' analogues, where the system reaches a
jammed state in which the lattice or space cannot be filled completely
\cite{adsorption}.

In the present paper we investigate the static pairwise annihilation
problem on complex networks, generalizing previous results obtained
for ordered lattices \cite{kenkre1,kenkre2}.  As such, the
annihilation problem on complex networks is interesting in its own
right since it provides a dynamical probe of the topological
properties of the structure on which it occurs.  These properties will
be different than, and complementary to, those associated with other
physical processes in complex networks, such as random walks
\cite{randomwalk}, Ising spin dynamics \cite{isingSW1}, target
annihilation \cite{target1,target2}, majority voter models and neural
network dynamics \cite{majority,laguna}, epidemic spreading
\cite{kuperman,pastor}, trapping \cite{trapping1,trapping2}, etc.

It is precisely these complementary features which lead us to consider
static annihilation as a probe of different complex networks, such as
ordered lattices, Erd\"os-R\'enyi random networks (also called Poisson
networks), scale-free networks and small-world networks.  Small-world
networks are particularly interesting, since they can be considered as
a class of networks lying at some intermediate point between ordered
lattices and fully random networks. They possess a large degree of
local clustering, but a relatively small minimal path length
connecting nodes throughout the system.  On the other hand, scale-free
networks are characterized by the presence of ``hubs'' or elements
with an extremely high connectivity.  These hubs hold the network
together and produce a very small minimal path length. Complex
networks are currently under intense investigation as useful
structures on which to model a variety of network related phenomena,
including electrical power grids, the Internet, human social
relationships, proteomic and genetic interactions, to mention just a
few examples \cite{watts,strogatz,barabasi,newman,mendes}.  We are
interested here in exploring the effects that arise on the static
annihilation process from the different features associated with the
different network topologies mentioned above.

\begin{figure}[t]
\begin{center}
\scalebox{0.4}{\includegraphics{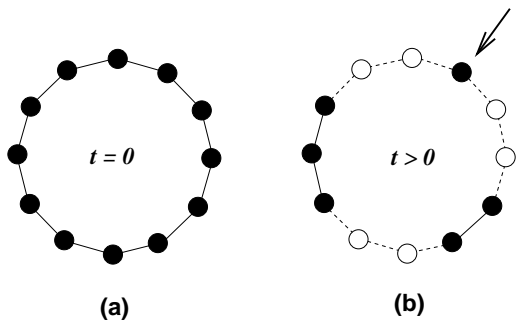}}
\end{center}
\caption{Schematic diagram illustrating the annihilation process in a
1D ring. (a) At time $t=0$ there are $N$ particles placed along the
ring (black circles).  (b) As time goes by, pairs of neighboring
particles annihilate at a constant rate $\alpha$. At some time $t>0$
some pairs have been annihilated (open circles). Note that all the
neighbors of the particle signaled with an arrow have been
annihilated. Since this particle has no neighbors left to annihilate
with, it will survive indefinitely.}
\end{figure}
\label{fig:ring}

Several exact results have previously been obtained for the static
annihilation problem on ordered lattices. Kenkre and Van Horn, e.g.,
considered a 1D ordered ring of $N$ sites, in which each site is
initially occupied by a particle, as in Fig.~\ref{fig:ring}a. At the
onset of the annihilation process, neighboring particles on the ring
begin to undergo mutual annihilation with rate constant $\alpha$. As
pairs of particles disappear from the ring, the possibility arises for
particles at certain sites to become isolated. Such particles end up
surviving indefinitely, with no neighbors left on either side to
annihilate with (see Fig.~\ref{fig:ring}b). For this model the authors
of Ref. \cite{kenkre1,kenkre2} were able to calculate exactly the
average fraction $f(t)$ of sites on the ring that remain populated at
time $t$. For an infinite 1D ring they found
\begin{equation}
f(t) = \exp\left(2e^{-\alpha t}-2\right),
\label{eq:kenkre1}
\end{equation}
from which the exact stationary survival fraction 
\begin{equation}
f_{\infty}=\lim_{t\to\infty} f(t)=e^{-2}
\label{eq:kenkre2}
\end{equation}
follows. A dynamical mean-field theory for higher dimensional ordered
structures was later given by Kenkre \cite{kenkre2}. In that work, the
long-time survival fraction on a square lattice in $D$ dimensions was
found to decrease as the dimensionality  $D$ (and the number
of potential annihilation partners) increases, specifically, via
\begin{equation}
f_\infty = (2D-1)^{D/(1-D)}.
\label{eq:kenkre3}
\end{equation}

Of course, on a translationally invariant structure the connectivity
of any site on the lattice is the same as that of any other. Each
site, moreover, has \emph{a priori} the same probability of ultimately
surviving the annihilation onslaught. It is interesting, therefore, to
see what effects are generated in the annihilation dynamics by
\emph{dispersion} in the local site coordination number. In a sense,
Poisson, scale-free and small-world networks are ideal structures on
which to explore these questions. Such structures allow, through a
simple variation of the network parameters, significant changes in the
average coordination number as well as a means to systematically
change the associated dispersion in local coordination. This is
especially true for scale-free networks, where the dispersion in the
local coordination number can become infinite.

In the present paper we generalize the results of
\cite{kenkre1,kenkre2} to a variety of complex networks that allow us
to explore these connectivity-related issues. First, in
Section~\ref{sec:mean-field} we present a mean-field theory of the
annihilation process on disordered networks with arbitrary
connectivity distributions. We then apply the general formalism to
three specific cases: (a) random networks in which every element has
exactly $K$ connections (Kronecker networks); (b) Erd\"os-R\'enyi
(i.e. Poisson) networks with average connectivity $K$; and (c)
scale-free networks. It is worth mentioning that the mean-field
approach requires the statistical independence of the network
elements. Thus, our formalism applies neither to ordered lattices nor
to small-world networks, where loops of connected sites are commonly
found.  Such loops generate non-trivial correlations between the
elements which are not taken into account in the mean-field
approach. However, case (a) above can be considered as an
approximation to ordered lattices with coordination number $K$. The
approximation becomes exact in the limit $K\to\infty$. Additionally,
as we show through numerical simulation in
Section~\ref{sec:small-world}, the mean-field approach qualitatively
describes the annihilation dynamics in ordered and small-world
structures.  We end this work in Section~\ref{sec:conclusions} with a
summary and discussion of our results.

\section{Mean-field theory}
\label{sec:mean-field}

\begin{figure}[t]
\begin{center}
\scalebox{0.4}{\includegraphics{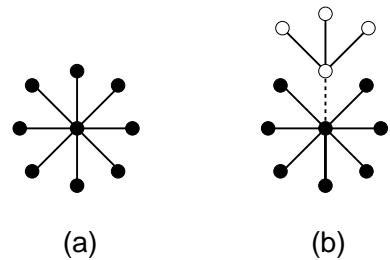}}
\end{center}
\caption{(a) A $k$-site (at the center) with $k$ connecting sites.
$k=8$ in this diagram.  (b) One of the connecting sites of the
$k$-site is an $m$-site with $m$ connections ($m = 4$ in the diagram).
The probability that an arbitrary connection (dashed line) of the
$k$-site goes to an $m$-site is $m f(m,t)/\sum_{m=0}^\infty mf(m,t)$.}
\label{fig:network}
\end{figure}

In this section we develop a simple mean-field theory of particle
annihilation in networks with an arbitrary connectivity distribution
$P(k)$.  This theory is exact as long as the connecting sites of each
element in the network are randomly chosen from anywhere in the
system.  In such a case, the correlations that arise due to the
occurrence of loops of connected sites can be neglected. The above is
not true in general either for small-world networks or for ordered
lattices. However, the mean-field approach presented here will allow
us to gain insight into the effects that the overall connectivity of
the system has on the annihilation dynamics. As we will see, since the
infinite 1D ordered lattice with nearest-neighbor connections is
loopless, our predictions recover the exact solution for this
structure.

We are interested in evaluating the fraction of surviving sites $f(t)$
at time $t$ in a network in which dimers (i.e. pairs of connected
sites) have been removed randomly.  We assume that the network is
initially full, and that the probability that a site has $k$
connections is given by $P(k)$.

We denote by $f(k,t)$ the fraction of surviving sites that are
connected to $k$ surviving sites at time $t$. We will refer to these
as $k$-sites. It is clear that $f(k,t=0)=P(k)$. Next, we determine the
time evolution of these densities by considering the three different
ways in which $f(k,t)$ can change:

\begin{enumerate}

\item As we remove dimers, a $k$-site can become a $(k-1)$-site if one
of its connecting sites is removed. To determine the rate at which
this happens we focus on one of the $k$ connections of the $k$-site
(dashed line in Fig.~\ref{fig:network}b). The probability that this
connection goes to an $m$-site is proportional to $m f(m,t)$
\footnote{The factor $m$ stands for the fact that it is more probable
to be connected to sites with a large number of connections than to
sites with a small number of connections.}. We have to normalize this
probability by dividing between $\sum_{m=1}^\infty m f(m,t)$ since the
fractions $f(m,t)$ change in time and they do not add up to 1 (except
at $t=0$).  Therefore, the chosen connecting site in
Fig.~\ref{fig:network}b will be an $m$-site with probability $m
f(m,t)/\sum_{m=1}^\infty m f(m,t)$. The $m$-site will be removed if
the dimer formed by this site and any of the other $(m-1)$ sites to
which it is connected (not considering the $k$-site we started from)
is removed. Adding over all $m$ and considering the $k$ connections of
the $k$-site, the total rate at which $k$-sites become $(k-1)$-sites
is
\[
k f(k,t)\frac{\sum_{m=0}^\infty m(m-1) f(m,t)}
{\sum_{m=0}^\infty m f(m,t)}.
\]

\item A $(k+1)$-site can become a $k$-site. For the same reasons as
before, this happens with probability
\[
 (k+1) f(k+1,t)\frac{\sum_{m=0}^\infty
m(m-1) f(m,t)} {\sum_{m=0}^\infty m f(m,t)}.
\]

\item Finally, $k$-sites may also be removed if any of the $k$ different 
dimers to which they belong is chosen for removal, which occurs with
probability $k f(k,t)$.
\end{enumerate}

Adding the three quantities above (with the correct sign) we find that
the densities $f(k,t)$ satisfy the following set of evolution
equations:

\begin{eqnarray}
\frac{\partial f(k,t)}{\partial t} &=&\nonumber 
\big[(k+1) f(k+1,t)- k f(k,t)\big]\times\nonumber\\
& &\frac{\sum\limits_{m=0}^\infty m(m-1) f(m,t)}
{\sum\limits_{m=0}^\infty m f(m,t)} - k f(k,t). 
\label{eq:for-f}
\end{eqnarray}
The above equations can be treated by generating functions as follows.
Let $F(z,t)$ be the generating function defined as
\begin{equation}
F(z,t)=\sum_{k=0}^\infty z^k f(k,t).
\label{eq:definition-F}
\end{equation}
Note that $f(t)=F(1,t)=\sum_{k=0}^\infty f(k,t)$ is just the overall
fraction of surviving elements at time $t$.  From Eq.~(\ref{eq:for-f})
it follows that $F(z,t)$ satisfies the equation
\begin{equation}
\frac{\partial F(z,t)}{\partial t}=\left[ (1-z)g(t)-z\right]
\frac{\partial F(z,t)}{\partial z},
\label{eq:for-F}
\end{equation}
where
\begin{equation}
g(t)=\frac{\sum\limits_{m=0}^\infty m(m-1) f(m,t)}
{\sum\limits_{m=0}^\infty m f(m,t)} =
\left[\frac{\left(\frac{\partial^2 F(z,t)}{\partial z^2}\right)}
{\left(\frac{\partial F(z,t)}{\partial z}\right)}\right]_{z=1}.
\label{eq:consistency-g}
\end{equation}
Equation (\ref{eq:for-F}) can be solved by the method of characteristics,
which gives:
\begin{equation}
F(z,t) = Q\left( 1 + (z-1) e^{-G(t)} - 
\int_0^t e^{-G(\tau)} d\tau\right),
\label{eq:sol-F}
\end{equation}
where
\begin{equation}
G(t) = t + \int_0^t g(\tau) d\tau,
\label{eq:def-G}
\end{equation}
and 
\begin{equation}
Q(z)=\sum_{k=0}^\infty z^k P(k).
\label{eq:def-Q}
\end{equation}
Note that $Q(z)$ is the generating function of the initial
distribution $P(k)$. It is also important to note that $g(t)$ must be
determined through the consistency equation (\ref{eq:consistency-g}).

It is clear that at the end of the annihilation process, only isolated
elements with $k=0$ will survive. Therefore, the fractions $f(k,t)$ should
fulfill the condition
\[
\lim_{t\to\infty}f(k,t)=0 \ \ \ \mbox{for all}\ \ \ k\geq1.
\]
The fraction $f_\infty$ of surviving elements at the end of
the annihilation process is then given by
\begin{equation}
f_\infty = \lim_{t\to\infty}f(0,t) = \lim_{t\to\infty}F(1,t).
\label{eq:f-infty}
\end{equation}

In what follows we apply the general formalism presented in this
section to obtain $f_\infty$ for networks with different topologies.

\subsection{Kronecker Delta Distribution}
\label{sec:kronecker}

We now apply the preceding formalism to the specific case in which
every site of the initial network has exactly $K$
connections. Therefore, the initial distribution $P(k)$ is given by a
Kronecker delta function: $P(k)=\delta_{k,K}$. This case can be
considered as an approximation to an ordered lattice with coordination
number $K$. Such an approximation becomes exact for $K\to\infty$.
However this case is particularly interesting when $K=2$ because,
given that in the ``thermodynamic'' limit there are no loops of
connected sites in the network, the system corresponds exactly to the
infinite 1D lattice with nearest-neighbor connections, for which the
exact solution is known (see Eqs.~(\ref{eq:kenkre1}) and
(\ref{eq:kenkre2}) and Refs.  \cite{kenkre1,kenkre2}). For this choice
of the initial distribution, Eq.~(\ref{eq:def-Q}) becomes
\[
Q(z) = \sum\limits_{k=0}^\infty z^k P(k) = z^K.
\]
From this result we find that the generating function for the
distribution of $k$-sites, Eq.~(\ref{eq:sol-F}), is given by
\begin{equation}
F(z,t)= \left[ 1 + (z-1) e^{-G(t)} - 
\int_0^t e^{-G(\tau)} d\tau\right]^K,
\label{eq:F-kro}
\end{equation}
and the consistency equation (\ref{eq:consistency-g}) becomes
\begin{equation}
g(t) = -(K-1)\frac{d}{dt}
\ln\left(1-\int_0^t e^{-G(\tau)} d\tau \right).
\label{eq:g-kro}
\end{equation}

To solve the  preceding integral equation we first integrate it from 0 to $t$,
which gives
\[
\int_0^t g(\tau)d\tau = -(K-1)\ln\left[1-\int_0^t e^{-G(\tau)}d\tau\right],
\]
or, equivalently,
\[
\exp\left[-\int_0^t g(\tau)d\tau\right]
 = \left(1-\int_0^t e^{-G(\tau)}d\tau\right)^{K-1}.
\]
Multiplying both sides of the above equation by $e^{-t}$ we obtain
\begin{equation}
e^{-G(t)} = e^{-t}\left(1-\int_0^t e^{-G(\tau)}d\tau\right)^{K-1},
\label{eq:interG}
\end{equation}
where we have used the definition of $G(t)$ given in
Eq.~(\ref{eq:def-G}).  The following change of variable will be very
useful to solve the integral equations:
\begin{equation}
u(t) = 1-\int_0^t e^{-G(\tau)}d\tau.
\label{eq:def-u}
\end{equation}
With this change of variable, Eq.~(\ref{eq:interG}) can be written as
\begin{equation}
\frac{d u(t)}{dt} = -e^{-t}\left[u(t)\right]^{K-1}.
\label{eq:for-u}
\end{equation}

The above equation can be solved easily. However, care must be taken
for the case $K=2$, which must be evaluated separately.  Assuming
$K>2$ first, the solution of Eq.~(\ref{eq:for-u}) with the initial
condition $u(0)=1$ is
\begin{equation}
u(t) = \left[1+(K-2)(1-e^{-t})\right]^{1/(2-K)}.
\label{eq:sol-u-kro}
\end{equation}

Although the complete hierarchy of distributions $f(k,t)$ can be
obtained from the preceding results, at this point we are interested
in the final fraction of surviving sites $f_\infty$. This can be
readily computed by noting from Eqs.~(\ref{eq:F-kro}) and
(\ref{eq:def-u}) that the overall fraction $f(t)=F(1,t)$ of surviving
sites at time $t$ is
\[
f(t) = [u(t)]^K = \left[1+(K-2)(1-e^{-t})\right]^{K/(2-K)}.
\]
Taking the limit $t\to\infty$ in the above equation we obtain
\begin{equation}
f_\infty = (K-1)^{K/(2-K)}\ \ \ (K > 2).
\label{eq:sol-finftykro}
\end{equation}
Since a square lattice in $D$ dimensions has coordination number
$K=2D$, the above result coincides with Eq.~(\ref{eq:kenkre3}), which
was first reported in Ref.~\cite{kenkre2}.

As mentioned before, the case $K=2$ has to be evaluated
separately. For $K=2$, the solution of Eq.~(\ref{eq:for-u}) with the
initial condition $u(0)=1$ is
\begin{equation}
u(t)=\exp\left(e^{-t}-1\right).
\label{eq:sol-u-kro-2}
\end{equation}
Therefore, the overall fraction $f(t)$ of surviving sites in this case
becomes
\[
f(t) = \left[u(t)\right]^2 = \exp\left(2e^{-t} -2\right),
\]
which, in the limit $t\to\infty$, gives
\begin{equation}
f_\infty = e^{-2} \ \ \ (K = 2).
\label{eq:f-kro-K2}
\end{equation}
This is the exact result for the 1D lattice with nearest-neighbor
connections. (See Eq. (\ref{eq:kenkre2}) and
Refs. \cite{kenkre1,kenkre2}.)

\begin{figure}[t]
\begin{center}
\scalebox{0.3}{\includegraphics{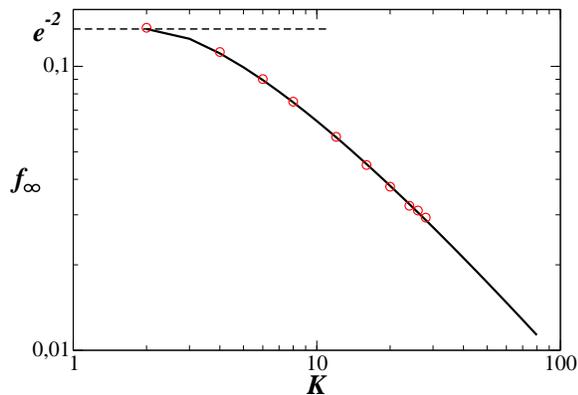}}
\end{center}
\caption{Fraction of surviving elements $f_\infty$ as a function of
the average network connectivity $K$ for networks with Kronecker
distribution.  The circles are the plot of the simulation data whereas
the solid curve is the graph of the theoretical prediction given in
Eqs.~(\ref{eq:sol-finftykro}) and (\ref{eq:f-kro-K2}). The dashed line
indicates the value $e^{-2}$ corresponding to the 1D ring with
nearest-neighbor interactions. To generate the simulation data we used
networks with $N=10^3$ sites. Each point is the average over 100
realizations.}
\label{fig:kronecker}
\end{figure}

Fig.~\ref{fig:kronecker} shows $f_\infty$ as a function of $K$.  The
solid curve is the theoretical result as predicted in
Eqs.~(\ref{eq:sol-finftykro}) and (\ref{eq:f-kro-K2}), whereas the
circles correspond to the computer simulation data.  Note from
Eq.~(\ref{eq:sol-finftykro}) that $f_\infty \sim 1/K$ for large values
of $K$, and therefore
\[
\lim_{K\to\infty}f_\infty = 0.
\]
So, the more connected the initial network, the smaller the fraction
of surviving sites at the end of the annihilation process.

For the sake of completeness, we compute the explicit form of the
distributions $f(k,t)$ for the case $K>2$. To this end, we rewrite
Eqs.~(\ref{eq:def-u}) and (\ref{eq:for-u}) as
\begin{eqnarray*}
e^{-G(t)} &=& e^{-t}\left[u(t)\right]^{K-1},\\
\int_0^t e^{-G(\tau)}d\tau &=& 1-u(t).
\end{eqnarray*}
Substituting the above expressions into Eq.~(\ref{eq:F-kro}) we obtain
for the generating function
\[
F(z,t) = \left[(z-1)e^{-t}[u(t)]^{K-1} + u(t)\right]^K,
\]
which, after expanding in powers of $z$ using the binomial theorem,
transforms into
\[
F(z,t) = \sum_{k=0}^K \binom{K}{k}\frac{\left(u(t)-e^{-t}[u(t)]^{K-1}\right)^K}
{\left(e^t[u(t)]^{2-K} -1\right)^k} z^k.
\]

From this result it follows immediately that the distribution $f(k,t)$
of surviving sites with $k$ connections at time $t$ is given by
\begin{equation}
f(k,t) = \binom{K}{k}\frac{\left(u(t)-e^{-t}[u(t)]^{K-1}\right)^K}
{\left(e^t[u(t)]^{2-K} -1\right)^k},
\label{eq:sol-f-kro}
\end{equation}
where for $u(t)$ we have to use Eq.~(\ref{eq:sol-u-kro}) if $K>2$, or
Eq.~(\ref{eq:sol-u-kro-2}) if $K=2$. In any case, from
Eq.~(\ref{eq:sol-f-kro}) we can see that $\lim_{t\to\infty}f(k,t) = 0$
for all $k\geq1$. Namely, at the end of the annihilation process, only
isolated sites with $k=0$ survive, as expected. However, for $k\geq1$
the rate at which $f(k,t)\to0$ as $t\to\infty$ depends on the value of
$k$. To see this, consider the distributions $f(k,t)$ and $f(k',t)$
with $k > k'$. Then, from Eq.~(\ref{eq:sol-f-kro}) we get
\[
\lim_{t\to\infty}\frac{f(k,t)}{f(k',t)} \sim\lim_{t\to\infty}
\frac{1}{\left(e^t[u(t)]^{2-K}-1\right)^{k-k'}} = 0.
\]
This result shows that  more connected sites annihilate faster than
less connected ones, as expected.

\subsection{Poisson Distribution}
\label{sec:poisson}

Next we treat the case in which the initial distribution of
connections is Poisson with mean $K$: $P(k) = e^{-K} K^k/ k!$. This
case corresponds to the Erd\"os-R\'enyi topology, for which $Q(z)$
becomes
\begin{equation}
Q(z) = \sum_{k=0}^\infty z^kP(k) = e^{-K}\sum_{k=0}^\infty z^k
\frac{K^k}{k!}  = e^{-K(1-z)}.
\end{equation}
Using this result in Eq.~(\ref{eq:sol-F}) we obtain
\begin{equation}
F(z,t)= \exp\left[-K\left((1-z)e^{-G(t)} + 
\int_0^t e^{-G(\tau)} ~ d\tau\right)\right],
\label{eq:Fpois}
\end{equation}
and the consistency equation (\ref{eq:consistency-g}) in this case is
\begin{equation}
g(t) = 
K \exp\left\{-t-\int_0^t g(\tau)~d\tau\right\} = K e^{-G(t)}.
\label{eq:g-pois}
\end{equation}
The above integral equation can be easily solved for $g(t)$, obtaining
\begin{equation}
g(t) = \frac{Ke^{-t}}{1+K(1-e^{-t})}.
\label{eq:sol-gpois}
\end{equation}
From Eq.~(\ref{eq:g-pois}) we have $e^{-G(t)}=g(t)/K$, from which it
follows that
\begin{eqnarray*}
e^{-G(t)} &=& \frac{e^{-t}}{1+K(1-e^{-t})}, \\
\int_0^t e^{-G(\tau)}d\tau &=& \frac{1}{K}\ln\left\{1+K(1-e^{-t})\right\}.
\end{eqnarray*}
Substituting the above results into Eq.~(\ref{eq:Fpois}) we obtain
\begin{eqnarray}
F(z,t) &=& \frac{e^{-g(t)}}{1+K(1-e^{-t})}\ e^{g(t)z}\label{eq:sol-Fpois1}\\
&=&\frac{e^{-g(t)}}{1+K(1-e^{-t})}\sum_{k=0}^\infty\frac{[g(t)]^k}{k!} z^k.
\label{eq:sol-Fpois2}
\end{eqnarray}
where we have used the fact that $g(t)=Ke^{-G(t)}$.  It follows
immediately from Eq.~(\ref{eq:sol-Fpois2}) that the
distribution $f(k,t)$ of $k$-sites at time $t$ is
\begin{equation}
f(k,t) = \frac{e^{-g(t)}}{1+K(1-e^{-t})}\frac{[g(t)]^k}{k!}.
\label{eq:sol-fpois}
\end{equation}

An interesting aspect of this result is the fact that the distribution
of $k$-sites continues to be Poissonian (though not normalized)
throughout the evolution of the system. Note also that, since
$\lim_{t\to\infty} g(t) = 0$ (see Eq.~(\ref{eq:sol-gpois})), it
follows from Eq.~(\ref{eq:sol-fpois}) that $\lim_{t\to\infty} f(k,t) =
0$ for all $k\geq1$, as expected. Again, the rate at which
this happens depends on the value of the connectivity
$k$. Indeed, considering the distributions $f(k,t)$ and $f(k',t)$,
with $k>k'$, from Eq.~(\ref{eq:sol-fpois}) and the fact that
$g(t)\to0$ as $t\to\infty$, we obtain
\[
\lim_{t\to\infty}\frac{f(k,t)}{f(k',t)} \sim \lim_{t\to\infty}
\left[g(t)\right]^{k-k'} = 0.
\]
Therefore, the higher the connectivity of a given site, the faster
it annihilates. 

The overall fraction $f(t)$ of surviving sites at time $t$, regardless
of their connectivity, is obtained from Eq.~(\ref{eq:sol-Fpois1}) with
$z=1$, which gives
\[
f(t) = F(1,t) = \frac{1}{1+K(1-e^{-t})}.
\]
Finally, the fraction $f_\infty$ of surviving sites at the end of the
annihilation process can be obtained from the above equation (or from
Eq.~(\ref{eq:sol-fpois}) with $k=0$) by taking the limit $t\to\infty$.
We obtain
\begin{equation}
f_\infty = \frac{1}{1+K}.
\label{eq:sol-finftypois}
\end{equation}
%

\begin{figure}[t]
\begin{center}
\scalebox{0.3}{\includegraphics{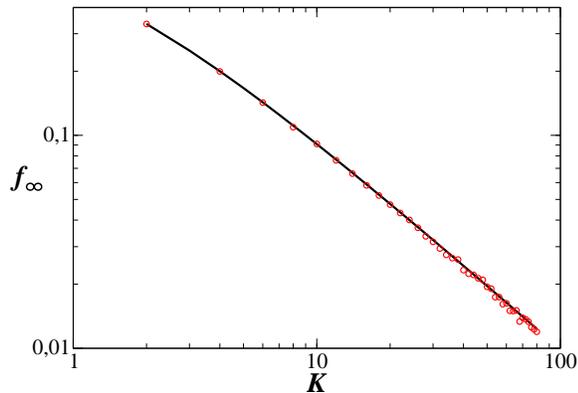}}
\end{center}
\caption{Fraction of surviving elements $f_\infty$ as a function of
the average network connectivity $K$ for Erd\"os-R\'enyi networks with
Poisson distribution.  The circles are the plot of the simulation data
whereas the solid curve is the graph of the theoretical prediction given
in Eq.~(\ref{eq:sol-finftypois}). Each point of the simulation data is
the average over 100 realizations using networks with $N=10^3$ sites.}
\label{fig:pois}
\end{figure}

Fig.~\ref{fig:pois} shows the graph of $f_\infty$ as a function of
$K$.  The solid curve is the plot of the theoretical prediction given
in Eq.~(\ref{eq:sol-finftypois}) and the circles correspond to the
simulation data. Note that, once again,
\[
\lim_{K\to\infty} f_\infty =0,
\] 
as for the Kronecker-delta case.

\subsection{Scale-free distribution}
\label{sec:scale-free}

In this section we consider scale-free networks for which
the initial distribution of connections acquires the form
\[
P(k)=\frac{1}{\zeta(\gamma)}k^{-\gamma},
\]
where $\zeta(\gamma)=\sum_{k=1}^\infty k^{-\gamma}$ is the Riemann zeta 
function. In this case, the generating function (\ref{eq:def-Q}) of the
initial distribution becomes
\begin{eqnarray}
Q(z) &=& \sum_{k=1}^\infty z^k P(k) = 
\frac{1}{\zeta(\gamma)}\sum_{k=1}^\infty\frac{z^k}{k^\gamma} \nonumber \\
&=& \frac{1}{\zeta(\gamma)} \mbox{Li}_\gamma(z),
\label{eq:Q-sf}
\end{eqnarray}
where $\mbox{Li}_\gamma(z)=\sum_{k=1}^\infty z^k/k^{\gamma}$ is the
polylogarithm function of order $\gamma$. This function has the
following properties, which will be of use for further analysis:
\begin{subequations}
\label{eq:Li-prop}
\begin{eqnarray}
\mbox{Li}_\gamma(0) &=& 0;\\
\mbox{Li}_\gamma(1) &=& \zeta(\gamma);\\
\frac{\partial \mbox{Li}_\gamma(z)}{\partial z} &=&
\frac{1}{z}\mbox{Li}_{\gamma-1}(z);\\
\frac{\partial^2 \mbox{Li}_\gamma(z)}{\partial z^2} &=&
\frac{1}{z^2}\left[\mbox{Li}_{\gamma-2}(z)-\mbox{Li}_{\gamma-1}(z)\right].\ \ 
\end{eqnarray}
\end{subequations}
The generating function $F(z,t)$ for this case is
\begin{equation}
F(z,t) = \frac{1}{\zeta(\gamma)}\mbox{Li}_\gamma
\left( 1 + (z-1) e^{-G(t)} - 
\int_0^t e^{-G(\tau)} d\tau\right),
\label{eq:F-sf1}
\end{equation}
and, using properties (\ref{eq:Li-prop}c) and (\ref{eq:Li-prop}d), the
consistency equation (\ref{eq:consistency-g}) becomes
\begin{equation}
g(t) = \frac{e^{-G(t)}}{u(t)}\frac{\mbox{Li}_{\gamma-2}(u(t))-
\mbox{Li}_{\gamma-1}(u(t))}{\mbox{Li}_{\gamma-1}(u(t))},
\label{eq:g-sf1}
\end{equation}
where $u(t)$ is defined as in Eq.~(\ref{eq:def-u}). Note from
Eqs.~(\ref{eq:F-sf1}) and (\ref{eq:def-u}) that the overall fraction
$f(t)$ of surviving particles at time $t$ is given by
\begin{equation}
f(t) = F(1,t) =  \frac{1}{\zeta(\gamma)}
\mbox{Li}_\gamma\left(u(t)\right).
\label{eq:f-sf1}
\end{equation}
Therefore, to compute $f(t)$ we first have to know $u(t)$. In order to
do so, we rewrite the consistency equation (\ref{eq:g-sf1}) as
\[
g(t) = \frac{d}{dt}
\left[\ln(u(t))-\ln\left(\mbox{Li}_{\gamma-1}(u(t))\right)\right],
\]
where we have used the fact that $du(t)/dt=-e^{-G(t)}$.
The last equation can be integrated from $0$ to $t$ taking
into account that $u(0) = 1$, which leads to
\[
\int_0^t g(t)dt = \ln\left(\frac{u(t)}{\mbox{Li}_{\gamma-1}(u(t))}\right)
+\ln\left(\zeta(\gamma-1)\right).
\]
Adding $t$ on both sides of the previous equation and exponentiating
we get
\[
e^{-G(t)} = \frac{\mbox{Li}_{\gamma-1}(u(t))}{u(t)},
\frac{e^{-t}}{\zeta(\gamma-1)}
\]
or equivalently,
\begin{equation}
\frac{u(t)}{\mbox{Li}_{\gamma-1}(u(t))}\frac{du(t)}{dt} =
-\frac{e^{-t}}{\zeta(\gamma-1)}.
\label{eq:for-u-sf}
\end{equation}
We have been so far unable to find an exact analytic solution to the
above nonlinear differential equation. However, we can solve it
numerically to obtain the asymptotic value
$u_\infty=\lim_{t\to\infty}u(t)$ with any desired accuracy. Once
$u_\infty$ has been determined, it follows from Eq.~(\ref{eq:f-sf1})
that the final fraction $f_\infty$ of surviving particles at the end
of the annihilation process will be given by
\begin{equation}
f_\infty = \frac{1}{\zeta(\gamma)}\mbox{Li}_\gamma(u_\infty).
\label{eq:f-sf2}
\end{equation}

In order to determine $u_\infty$, we first integrate Eq.~(\ref{eq:for-u-sf})
from $0$ to $t$, which gives
\[
\int_{u(t)}^1 \frac{u'}{\mbox{Li}_{\gamma-1}(u')}du' =
\frac{1}{\zeta(\gamma-1)}\left(1-e^{-t}\right).
\]
Taking the limit $t\to\infty$ in the previous equation we obtain
\begin{equation}
\int_{u_\infty}^1 \frac{u'}{\mbox{Li}_{\gamma-1}(u')}du' =
\frac{1}{\zeta(\gamma-1)}.
\label{eq:int-for-u}
\end{equation}
An important conclusion can immediately be drawn from this equation.
Note that, since $\zeta(\gamma-1)=\infty$ for $1<\gamma\leq2$, the
above equation in this range becomes
\[
\int_{u_\infty}^1 \frac{u'}{\mbox{Li}_{\gamma-1}(u')}du' =0\ \ \ 
\mbox{ for } \ \ \ 1<\gamma\leq2,
\]
which has the solution $u_\infty = 1$. From Eq.~(\ref{eq:f-sf2}) and
property (\ref{eq:Li-prop}b), we obtain
\[
f_\infty = 1  \ \ \ \mbox{ for }\ \ \ 1 < \gamma \leq2.
\]
In other words, \emph{essentially all the particles of a scale-free
network with infinite average connectivity survive the annihilation
process}.  This result is easy to interpret. What holds a scale-free
network together are the ``hubs'', i.e., the sites with an extremely
high number of connections \cite{target1}. But the hubs represent a
negligible fraction of the total number of sites in the
network. However, a randomly chosen site will be connected to a hub
with certainty.  Therefore, in this range of values of $\gamma$, the
hubs are annihilated from the very beginning of the process. But once
the hubs have been annihilated, the network disintegrates and only
isolated sites are left. Hence, $f_\infty = 1$ in the thermodynamic
limit.

\begin{figure}[t]
\begin{center}
\scalebox{0.4}{\includegraphics{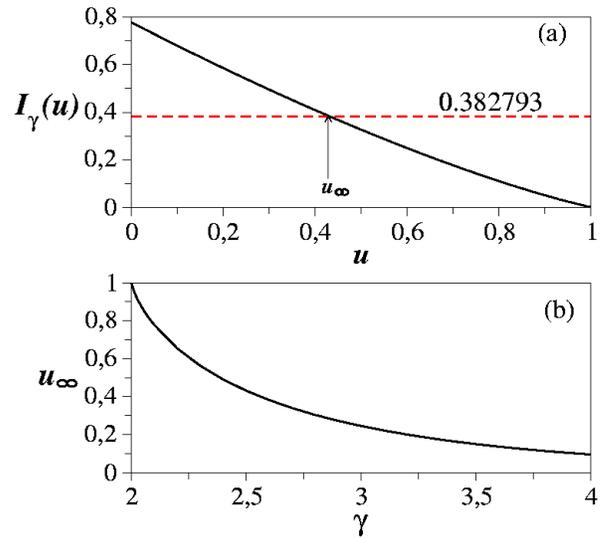}}
\end{center}
\caption{Features in scale-free networks. (a) The solid curve is the
graph of $I_\gamma(u)$ (defined in Eq.~(\ref{eq:def-I})) as a function
of $u$ for the case $\gamma=2.5$. The dashed line indicates the value
$1/\zeta(1.5)\approx 0.382793$, whose intersection with $I_\gamma(u)$
in this case gives $u_\infty\approx0.431$.  (b) $u_\infty$ as a
function of $\gamma$. This curve was obtained by solving numerically
the transcendental equation (\ref{eq:trascendental}).}
\label{fig:raices}
\end{figure}

\begin{figure}[t]
\begin{center}
\scalebox{0.4}{\includegraphics{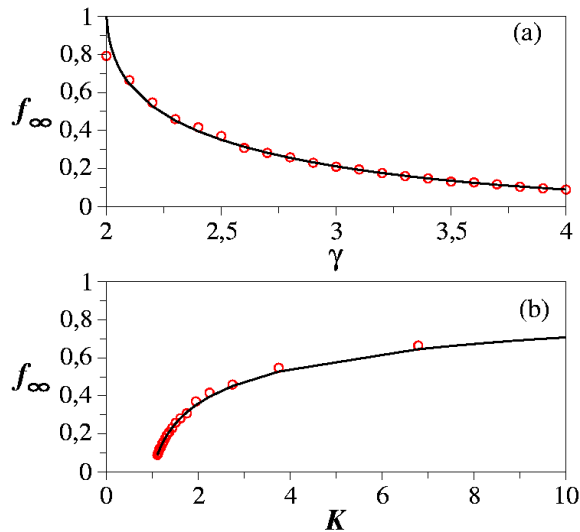}}
\end{center}
\caption{(a) Surviving fraction $f_\infty$ in scale-free networks as a
function of $\gamma$.  The solid curve corresponds to the theoretical
result as predicted by Eqs.~(\ref{eq:f-sf2}) and (\ref{eq:int-for-u}),
whereas the circles are the data obtained from computer
simulation. The simulation was carried out for networks with $N=10^4$
sites. Each point is the average over 100 realizations. (b) Same data
as before but plotted against the average network connectivity
$K=\zeta(\gamma-1)/\zeta(\gamma)$.}
\label{fig:scalefree}
\end{figure}

To obtain $f_\infty$ for $\gamma>2$  it is convenient to define the
function $I_\gamma(u)$ as
\begin{equation}
I_\gamma(u) = \int_u^1\frac{u'}{\mbox{Li}_{\gamma-1}(u')} du'.
\label{eq:def-I}
\end{equation}
With this definition, Eq.~(\ref{eq:int-for-u}) can be written as
\begin{equation}
I_\gamma(u_\infty) = \frac{1}{\zeta(\gamma-1)}.
\label{eq:trascendental}
\end{equation} 
This is a transcendental equation for $u_\infty$. Note that
$I_\gamma(u)$ is a continuous monotonically decreasing function of
$u$. For example, the solid line in Fig.~\ref{fig:raices}a is the
graph of $I_\gamma(u)$ for $\gamma = 2.5$, whereas the dashed line
indicates the value of $1/\zeta(\gamma-1)$, which for $\gamma=2.5$ is
$1/\zeta(1.5)\approx0.382793$. Then, $u_\infty$ is the value of $u$ at
which the solid curve intersects the dashed one, which for the case
shown in Fig.~\ref{fig:raices}a, is $u_\infty \approx 0.431$. Since
$I_\gamma(u)$ is a ``well behaved'' function for every $\gamma$, we
can find the roots of Eq.~(\ref{eq:trascendental}) with any desired
accuracy.  Fig.~\ref{fig:raices}b shows $u_\infty$ as a function of
$\gamma$.  Using these results in Eq.~(\ref{eq:f-sf2}), we obtain the
data reported in Fig.~\ref{fig:scalefree}a, in which $f_\infty$ is
plotted as a function of $\gamma$. The solid curve depicts the
theoretical result and the circles are data from the simulation of the
system in the computer. The small difference between the theoretical
prediction and the simulation at $\gamma=2$ is due to finite size
effects in the computer simulation.

Finally, Fig.~\ref{fig:scalefree}b shows the graph of $f_\infty$
as a function of the average network connectivity $K$, which for
scale-free networks is given by $K=\zeta(\gamma-1)/\zeta(\gamma)$.  Note that
$f_\infty$ \emph{increases} with $K$ and asymptotically approaches its
maximum value $f_\infty=1$ as $K\to\infty$. As explained before,
due to the presence of ``hubs'' in scale-free networks, we have 
\[
\lim_{K\to\infty}f_\infty = 1,
\]
contrary to what happens in Kronecker and Poisson networks, for which
$\lim_{K\to\infty}f_\infty = 0$.

\section{Extension to Small-World Networks}
\label{sec:small-world}

Small-world networks pose a special challenge for analytical treatment
since we have to deal with both ordered short-range and random
long-range interactions simultaneously. Although some analytic results
have been obtained for percolation and ferromagnetic-like systems in
small-world networks \cite{percolation,isingSW2,isingSW3,isingSW4}, to
the best of our knowledge there does not yet exist a general formalism
to study the dynamical properties of this kind of networks given a
specific interaction rule between the elements. In particular, for the
annihilation problem we are considering, the ordered short-range
connections produce ``loops'' of connected sites, which in turn give
rise to nontrivial correlations between the elements. Such
correlations are not taken into account in the mean-field theory we
have presented. Therefore, we do not expect our formalism to be
exactly applicable to small-world networks.

However, since small-world networks can be considered as something
between ordered lattices and fully random networks, we do expect the
static annihilation process in SW structures to be similar to that in
Kronecker or Poisson networks. In this section we present results
obtained from the numerical simulation of the system on small-world
networks.  Such networks are constructed by starting with an ordered
ring of $N$ sites and connectivity $K$ in which each site is connected
to its $K/2$ nearest neighbors on the left and $K/2$ nearest neighbors
on the right. Disorder is then introduced into this ordered structure
by disconnecting with probability $p$ each bond in the lattice
emerging from a given site and reconnecting it to another randomly
chosen site to which it was not originally connected (provided that
the rewiring does not result in two or more disconnected subnetworks).

Thus, the reconnection or rewiring probability $p$ is a measure of the
disorder of the system, and allows the system to be tuned from a
completely ordered ($p=0$) to a completely disordered ($p=1$)
structure. It is also a direct measure of the number of ``short-cuts''
that exist in the network. Note that the rewiring algorithm leaves the
average connectivity $K$ associated with the original ordered
structure unchanged, but changes the dispersion in the local
connectivity. Indeed, after rewiring any given site may be connected
to a greater or fewer number of sites than before rewiring.
Therefore, the average connectivity and the magnitude of fluctuations
in the local connectivity are independently adjustable through the
parameters $K$ and $p$.

\begin{figure}[t]
\begin{center}
\scalebox{0.4}{\includegraphics{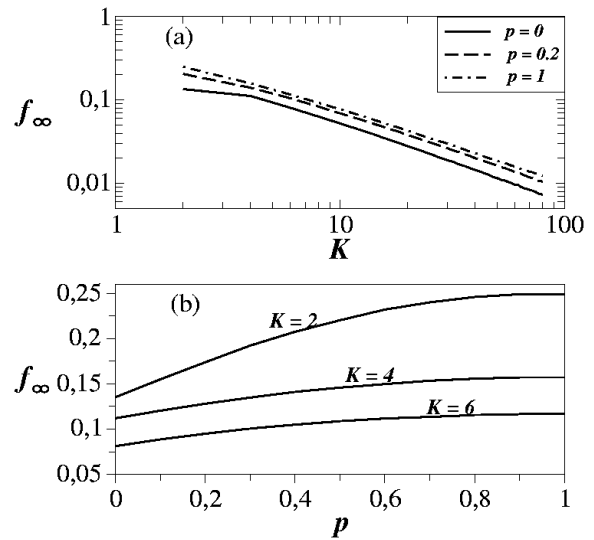}}
\end{center}
\caption{(a) Surviving fraction $f_\infty$ in small-world networks as
a function of $K$ for three values of the rewiring probability.  These
three values correspond to an ordered ring ($p=0$), a small-world
network with 20\% of shortcuts ($p=0.2)$, and a fully random network
($p=1$).  (b) Fraction of surviving elements $f_\infty$ as a function
of $p$ for three values of the average network connectivity $K$. In
both figures the data were computed numerically through computer
simulation using networks with $N=10^3$ sites. Each point is the
average over 1000 realizations.}
\label{fig:sw}
\end{figure}

Fig.~\ref{fig:sw}a shows $f_\infty$ as a function of $K$ for three
different values of the rewiring probability: $p=0$ (ordered ring),
$p=0.2$ (small-world network) and $p=1$ (fully random network).  Note
that $f_\infty$ behaves as $f_\infty\sim1/K$ for large values of $K$
regardless of the value of the rewiring probability $p$. Therefore,
small-world networks behave like Kronecker and Poisson networks.  On
the other hand, Fig.~\ref{fig:sw}b shows $f_\infty$ as a function of
$p$ for three values of the average network connectivity $K$.
Interestingly, the numerical results of this figure indicate that the
fraction $f_{\infty}$ of surviving particles \emph{increases} with an
increasing number of small-world shortcuts. One might think that such
shortcuts would enhance particle annihilation between sites very far
away, and thereby lead to a decrease in the survival
fraction. However, in order for a small-world network to achieve the
short path length with which it is associated, a significant number of
sites have to be more connected than others in the system. Such sites
act as gateways, allowing access across the system from any
direction. Since the average network connectivity is preserved during
rewiring, the existence of highly connected elements implies the
existence of many elements with a connectivity smaller than average.
Consequently, there will exist situations in which sites with high
connectivity are connected to many sites of below average
connectivity. As our mean-field formalism shows, a highly-connected
site is very likely to annihilate quickly with one of its many
partners, leaving the remaining sites to which it was connected more
isolated than they would have been if they had been part of an ordered
network. Hence, it appears that dispersion in the local connectivity
leads to enhanced survivability in the annihilation problem.

\section{Summary and discussion}
\label{sec:conclusions}

In this work he have analyzed the static annihilation process on
complex networks. In this problem, pairs of connected particles
annihilate at a constant rate.  Through a mean-field approach we were
able to compute exact general expressions for the fraction of
surviving particles with $k$ connections at time $t$ for random
networks with arbitrary connectivity distribution $P(k)$.  Our results
indicate that Kronecker and Poisson networks exhibit a similar
behavior in that the final fraction of surviving particles $f_\infty$
asymptotically behaves as $1/K$ for large values of $K$, the average
connectivity of the initial network. Numerical simulations show that
this is also true for ordered lattices and small-world networks.  So,
although the mean-field approach is not exactly applicable to ordered
latices and small-world networks, it predicts qualitatively the
correct asymptotic limit $f_\infty\sim1/K$ for such systems.

The situation is radically different in scale-free networks,
characterized by the connectivity distribution $P(k)\sim
k^{-\gamma}$. For this kind of network, our theory predicts that
$f_\infty =1$ for $1<\gamma\leq2$ and that
$\lim_{\gamma\to\infty}f_\infty = 0$. Thus, in the range
$1<\gamma\leq2$ in which the average network connectivity $K$
diverges, all the particles (or almost all in finite systems) survive
the annihilation process.  The above result is a consequence of the
fact that highly connected sites annihilate faster than poorly
connected ones. While this happens in every network, it has dramatic
consequences in scale-free structures in which, once the hubs are
annihilated, the network breaks apart and only isolated sites are
left. This kind of behavior is reminiscent of the breakdown of
scale-free networks when the highly connected sites are deliberately
attacked \cite{target1}.  Here we have shown that the annihilation of
the hubs does not only break the network apart, but that it pulverizes
the network.

Diffusion annihilation dynamics of the kind $A+A\to\emptyset$ has
recently been studied on complex networks using an analogous formalism
as the one presented in this work \cite{annihilation}. In diffusion
annihilation, the particles, rather than being fixed to the vertices,
are free to move (diffuse) throughout the network. Two particles
annihilate if they collide at some vertex of the network for the first
time. The authors of Ref. \cite{annihilation} show that the fraction
of surviving particles decreases in time as $t^{-1}$ for Poisson-like
networks and as $t^{-\beta}$ for scale-free networks, where $\beta$
depends on the scale-free exponent. Obviously, contrary to what
happens in the static case considered in this work, there are no
surviving particles in diffusion annihilation. Thus, our results
describe situations that are, in some sense, complementary to those
analyzed by Ref.~\cite{annihilation}.

\section*{ACKNOWLEDGMENTS}

M.F. Laguna, M. Aldana and P. E. Parris thank the hospitality of the
Consortium of the Americas for Interdisciplinary Science, University
of New Mexico at Albuquerque, U.S.A.  The authors thank G. Abramson,
M. Kuperman, M. Fuentes and F. Leyvraz for valuable discussions.  This
work was partially supported by NSF grants INT-0336343, DMR-0097204,
DMR-0097210, and DARPA-N00014-03-0900, as well as by DGAPA-UNAM grant
IN100803.


\begin{thebibliography}{9}

\bibitem {redner1} L. Frachebourg, P.L. Krapivsky, and S. Redner,
  Jour. Phys. A: Math. Gen.  {\bf 31}, 2791 (1998).

\bibitem{redner2} E. BenNaim, S. Redner, and P.L. Krapivsky,
  Jour. Phys. A: Math. Gen.  {\bf 29}, L561 (1996).
 
\bibitem{molec}P. Avakian and R. Merrifield, Mol. Cryst. {\bf 5}, 37
  (1968); N.E. Geacintov and C.E. Swenberg, in \textit{Organic
  Molecular Photophysics}, edited by J.B. Birks (Willey, New Tork,
  1973), Vol. I;  V.M. Kenkre, Exciton Dynamics in Molecular Crystals
  and Aggregates: the Master Equation Approach, in Springer Tracts in
  Modern Physics, Vol. 94, edited by G. Hoehler (Springer, Berlin,
  1982).

\bibitem{vicsek} T. Vicsek, A. Czir\'ok, E. Ben-Jacob, I. Cohen, and
  O. Shochet, Phys. Rev. Lett. {\bf 75}, 1226 (1995).

\bibitem{selfdriven} M. Aldana and C. Huepe, Jour. Stat. Phys.{\bf 112}, 
  135 (2003).

\bibitem{pyc1} J.R. Wilson and G.J. Mathews, Astrophysical Jour. {\bf
  610}, 368 (2004).

\bibitem{pyc2} D.G. Yakovlev and K.P. Levenfish, Cont. Plasma
  Phys.{\bf 43} 390 (2003).

\bibitem{pyc3} H. Kitamura, Astrophysical Jour. {\bf 539}, 888 (2000).

\bibitem{adsorption} J.W. Evans, Rev. Mod. Phys. {\bf 65}, 1281
  (1993).

\bibitem {kenkre1}V.M. Kenkre and H.M. Van Horn, Phys. Rev. A
  {\bf 23}, 3200 (1981).

\bibitem {kenkre2}V.M. Kenkre, Journal of Statistical Physics {\bf
  30}, 293 (1983).

\bibitem{randomwalk} J.D. Noh and H. Rieger, Phys. Rev. Lett. {\bf
  92}, 118701 (2004).

\bibitem{isingSW1}  C.P Herrero. Phys. Rev. E. {\bf 65}, 066110 (2002). 
 
\bibitem{target1} R. Albert, H. Jeong, and A.-L. Barab\'asi, Nature
  {\bf 406}, 378 (2000).

\bibitem{target2} F. Jasch and A. Blumen, Phys. Rev. E. {\bf 63},
  041108 (2001).

\bibitem{majority} M. Aldana and H. Larralde. Phys, Rev. E. {\bf 70}
  066130 (2004).
 
\bibitem{laguna}M.F. Laguna, G. Abramson and D.H. Zanette, Physica A
  {\bf 329}, 459 (2003); Complexity {\bf 9}, No. 4, 31, Willey
  Periodicals (2004).

\bibitem {kuperman}M. Kuperman and G. Abramson,
  Phys. Rev. Lett. {\bf 86}, 2909 (2001).

\bibitem{pastor} R. Pastor-Satorras and A.Vespignani,
  Phys. Rev. Lett. {\bf 86}, 3200 (2001).

\bibitem{trapping1} L.K. Gallos, Phys. Rev. E. {\bf 70}, 046116
  (2004).

\bibitem{trapping2} F. Jasch and A. Blumen, Phys. Rev. E. {\bf 64},
  066104 (2001).

\bibitem {watts}D.J. Watts and S.H. Strogatz, Nature (London)
  {\bf 393}, 440 (1998).

\bibitem{strogatz} S.H. Strogatz, Nature. {\bf 410}, 268 (2001).

\bibitem{barabasi} R. Albert and A.-L. Barab\'asi,
  Rev. Mod. Phys. {\bf 74}, 47 (2002).

\bibitem{newman} M.E.J. Newman, SIAM Rev. {\bf 45}, 167 (2003).

\bibitem{mendes} S.N. Dorogovtsev and J.F.F. Mendes, Adv. Phys. {\bf
  51}, 1079 (2002).

\bibitem{percolation} C. Moore and M.E.J Newman, Phys. Rev. E. {\bf
  62}, 7059 (2000).

\bibitem{isingSW2} J. Viana-Lopes, Yu.G. Pogorelov, J.M.B. Lopes dos
  Santos and R. Toral, cond-math/0402138.

\bibitem{isingSW3} R. Nikoletopoulos, A.C.C. Cohen, I. P\'erez
  Castillo, N.S. Skantzos, J.P.L. Hatchett and B. Wemmenhove,
  Jour. Phys. A: Math. Gen. {\bf 37}, 6455 (2004).

\bibitem{isingSW4} A. Barrat and M. Weigt, Eur. Phys. Jour. B. {\bf
  13}, 547 (2000).


\bibitem{annihilation} M. Catanzaro, M. Boguna, and
  R. Pastor-Satorras, cond-math/0407447, (2004).

\end{thebibliography}
\end{document}